\documentclass[a4paper]{PoS}
\usepackage[T1]{fontenc}
\usepackage{braket}
\usepackage{amsmath,amssymb,graphicx}
\usepackage{mcite}

\def\helacnlo{\textsc{Helac-Nlo}}
\def\amcnlo{a\textsc{Mc@Nlo}}
\def\mcnlo{\textsc{Mc@Nlo}}
\def\powheg{\textsc{Powheg}}
\def\powhegbox{\textsc{Powheg-Box}}
\def\deductor{\textsc{Deductor}}
\def\pyQ{\textsc{Pythia6Q}}
\def\pythia{\textsc{Pythia}}
\def\amcnlopyQ{\amcnlo+\pyQ}
\def\amcnlopy{\amcnlo+\pythia}
\def\powhegpy{\powheg+\pythia}
\def\heldeductor{\helacnlo+\deductor}

\title{$pp \to t\bar{t}j + X$ matched to the Nagy-Soper parton
shower at NLO QCD}

\ShortTitle{$pp \to t\bar{t}j + X$ matched to the Nagy-Soper parton
shower at NLO QCD}

\author{\speaker{Manfred Kraus}\thanks{Preprint number: TTK-15-21}\\
        Institute for Theoretical Particle Physics and Cosmology\\
        RWTH Aachen University \\
        D-52056 Aachen, Germany\\
        E-mail: \email{kraus@physik.rwth-aachen.de}}

\abstract{We briefly summarize the Nagy-Soper parton shower and the
 \mcnlo{}-like matching scheme. Results obtained using
 \helacnlo{} framework in conjunction with \deductor{}
 for top quark pair production in association with one hard jet at the
 LHC are presented. A comparison of our results with other matching
 schemes and other parton showers is also discussed for various
 observables.}

\FullConference{The European Physical Society Conference on High Energy Physics\\
		22--29 July 2015\\
		Vienna, Austria}

\begin{document}

\section{Introduction}
High energy experiments like Tevatron or LHC study
the Standard Model of particle physics and its possible extensions. At
the core of these experimental studies are Monte Carlo (MC) generators
that are based on factorization theorems. Their construction usually
involves additional approximations and phenomenological models.
However, this approach allows the simulation of particle scatterings
starting from the hard interaction and then dressing the external legs
with further radiation generated by the parton shower.
This will evolve the hard state down to a low energy scale where
non-perturbative effects are important. At that point
hadronization models are employed.

In order to keep up with the increasing precision of experimental data,
MC generators have to be improved. There are several options to
achieve this. For fixed-order calculations the inclusion of
next-to-leading order (NLO) corrections in quantum chromodynamics
(QCD) are widely automated~\cite{Bevilacqua:2011xh, *Cullen:2011ac,
 *Cullen:2014yla,*Cascioli:2011va, Alwall:2014hca}.
Currently, efforts are ongoing in the automatization of electroweak
corrections~\cite{Actis:2012qn,*Kallweit:2014xda} while
NNLO QCD corrections are only available for $2\to 2$ processes.

Fixed-order calculations usually suffer from large logarithms, which
can be resummed to all orders in perturbation theory using
analytical resummation~\cite{Cacciari:2011hy, *Kulesza:2015vda,
*Kidonakis:1998bk, *Wang:2014mqt} or parton shower methods
\cite{Gleisberg:2008ta,*Corcella:2002jc,Sjostrand:2007gs,
Sjostrand:2006za}.
Matching of NLO fixed-order calculations to parton showers has the
benefit of combining several features. Well separated partons are
then correctly described by matrix elements in perturbative quantum
field theory, whereas the soft and collinear parton splittings are
generated by parton showers. There are several methods in the
literature to do the matching, the most popular ones being
\mcnlo{}~\cite{Frixione:2002ik, *Frixione:2003ei} and \powheg{}
\cite{Nason:2004rx, *Frixione:2007vw}.
The current matching and merging schemes are all limited
by the accuracy of the shower algorithms, that include only
leading colour (LC) and leading logarithmic (LL) accuracies and no spin
correlations. In order to go beyond these approximations one has to
include soft-gluon interferences and other subleading effects.
In Ref.~\cite{Platzer:2012np} it has been shown that these subleading
effects can be sizable for specific observables.

In this proceeding, we first briefly review the basics of the
Nagy-Soper shower. In section 3 we explain the matching scheme
for this shower and in section 4 we show first results for the
$pp\to t\bar{t}j + X$ production at the LHC. Finally, we conclude and
give an outlook for future improvements of this work in section 5.

\section{Nagy-Soper parton shower}
Here, we give a very brief summary of the Nagy-Soper parton
shower introduced by Zoltan Nagy and Davison Soper~\cite{Nagy:2007ty,
Nagy:2012bt}. We only highlight the necessary concepts to understand the
parton shower matching, while a more thorough discussion can be found
in Ref.~\cite{Czakon:2015cla}.
We start from the all-order expression of the expectation value of
an observable $F$, for a $2 \to m$ process
\begin{equation}
  \sigma[F] = \sum_m \frac{1}{m!}\int
  [d\{p,f,s^\prime,c^\prime,s,c\}_m] F(\{p,f\}_m) \braket{\{s^\prime,
  c^\prime\}_m|\{s,c\}_m}\rho(\{p,f,s^\prime,c^\prime,s,c\}_m)\;,
  \label{FO_expectation}
\end{equation}
where $f_i, s_i, c_i$ and $p_i$ represent the flavour, spin, colour and
momentum of a particle. The generalized phase space integration
measure, $[d\{p,f,s^\prime,c^\prime,s,c\}_m]$, includes the
integration over the initial state momentum fractions $\eta_a$ and
$\eta_b$ as well as the summation over spin and colour indices.
The matrix element, $\ket{\mathcal{M}(\{p,f\}_m)}$, is a vector in
colour and spin space and its square can be rewritten in the form of
a quantum density matrix defined as
\begin{equation}
  \rho(\{p,f,s^\prime,c^\prime,s,c\}_m) =
  \mathcal{M}^*(\{p,f,s^\prime,c^\prime,s,c\}_m)\mathcal{M}
  (\{p,f,s^\prime,c^\prime,s,c\}_m)
  \frac{f_{a}(\eta_a,\mu_F^2)f_{b}(\eta_b,\mu_F^2)}
  {4n_c(a)n_c(b)\times flux}\;.
\end{equation}
The parton shower evolves this quantum density from the hard scale,
$t_0$, of the scattering process down to the low scale, $t_F$, where
non-perturbative physics models can be applied. The evolution of the
system is described by a unitary operator $U(t_F,t_0)$, which
satisfies
\begin{equation}
  \frac{dU(t,t_0)}{dt} = \left[\mathcal{H}_I(t)  -
  \mathcal{V}(t) \right]U(t,t_0)\;.
\end{equation}
Here the evolution operator $U(t_F,t_0)$ is given by the real
splitting operator, $\mathcal{H}_I(t)$, which describes the emission
of resolved particles and the virtual operator, $\mathcal{V}(t)$,
which describes unresolved emissions. The above differential
equation is solved by the following operator
\begin{equation}
  U(t_F,t_0) = N(t_F,t_0) + \int_{t_0}^{t_F} d\tau \left[
  \mathcal{H}_I(\tau) - \mathcal{V}_S(\tau)\right]U(\tau,t_0)\;,
\end{equation}
with the normal Sudakov form factor
\begin{equation}
  N(t_F,t_0) = \exp\left(- \int_{t_0}^{t_F}
  d\tau~\mathcal{V}_E(\tau) \right)\;.
\end{equation}
The virtual splitting operator is decomposed into
$\mathcal{V}(t)=\mathcal{V}_E(t)+\mathcal{V}_S(t)$, where
$\mathcal{V}_E(t)$ is the colour diagonal part and $\mathcal{V}_S(t)$
the colour off-diagonal part with subleading contributions. The colour
diagonal part will be exponentiated while the off-diagonal
contribution is treated as perturbation.

This shower concept differs in many ways from conventional showers
and has been partly implemented in a MC program \deductor{}
\cite{Nagy:2014mqa}. A main conceptual difference is that
the splitting functions are derived using factorization on the
amplitude level~\cite{Nagy:2007ty}. In addition, initial state
charm and bottom quarks are treated as massive and PDFs used in the
shower are evolved according to the shower splitting kernels
\cite{Nagy:2007ty,Nagy:2014oqa}. The shower is consistently able to
include spin and colour correlations throughout the whole parton
evolution \cite{Nagy:2012bt,Nagy:2008ns,Nagy:2008eq}. The ordering
parameter in the evolution is a virtuality based one and implements
the validity of the on-shell approximation in each step of the
evolution~\cite{Nagy:2014nqa}
\begin{equation}
  e^{-t_l} = \frac{\Lambda_l^2}{Q^2}\;, \qquad
  \Lambda_l^2 = \frac{|(\hat{p}_l\pm \hat{p}_{m+1})^2-m^2(l)|}
  {2p_l\cdot Q}Q^2\;.
\end{equation}
Finally, the Nagy-Soper shower uses a global momentum mapping that
improves resummation effects in Drell-Yan $Z$-production
\cite{Nagy:2009vg}.

\section{Parton shower matching}
At next-to-leading order the quantum density matrix has to be
extended by including the virtual and real matrix elements
\begin{equation}
  |\rho) = \underbrace{|\rho_m^{(0)})}_{\text{Born} , \;
  \mathcal{O}(1)} + \underbrace{|\rho_m^{(1)})}_{\text{Virtual},\
  \mathcal{O}(\alpha_s)} + \underbrace{
  |\rho_{m+1}^{(0)})}_{\text{Real}, \; \mathcal{O}(\alpha_s)} +
  \mathcal{O}(\alpha_s^2)\;.
\end{equation}
Applying the shower evolution operator to this density generates
a spurious non-zero contribution at $\mathcal{O}(\alpha_s)$, which
can be easily seen by expanding the evolution operator
\begin{equation}
  (F| U(t_F,t_0)|\rho) \approx (F|\rho) + \int_{t_0}^{t_F}d\tau~
   (F|\left[ \mathcal{H}_I(\tau) - \mathcal{V}(\tau)\right]
  |\rho_m^{(0)}) + \mathcal{O}(\alpha_s^2)\;.
\end{equation}
As a consequence, the total cross section is changed and the first
emission is double counted. In order to remove this additional
contribution the \mcnlo{} approach has been applied.
We will first focus on fully inclusive processes like
$pp \to t\bar{t}$ and then we will explain the matching of exclusive
processes which already suffer from divergences at the LO.
\subsection{Fully inclusive processes}
The idea is to redefine the quantum density matrix by providing
suitable counterterms for the shower contribution
\begin{equation}
  |\bar{\rho}) \equiv |\rho) - \int_{t_0}^{t_F} d\tau \left[
  \mathcal{H}_I (\tau) - \mathcal{V}(\tau)\right]|\rho_m^{(0)})+
  \mathcal{O}(\alpha_s^2)\;.
\end{equation}
By dropping the infrared cutoff $(t_F \to \infty)$ we observe that the
shower naturally incorporates the NLO subtraction scheme
\begin{equation}
\begin{split}
  &\int_{t_0}^\infty d\tau~\mathcal{H}_I(\tau) = \sum_l
  \mathbf{S}_l\int_{t_0}^\infty d\tau\delta(\tau-t_l)
  \Theta(\tau-t_0) = \sum_l\mathbf{S}_l\Theta(t_l-t_0)\;,\\
  &\int_{t_0}^\infty d\tau~\mathcal{V}(\tau) = \sum_l \int d\Gamma_l
  \mathbf{S}_l\Theta(t_l-t_0) \equiv \mathbf{I}(t_0) +
  \mathbf{K}(t_0)\;.
\end{split}
\end{equation}
Thus, we see that matching is a two-step procedure
\begin{equation}
\begin{split}
  \bar{\sigma}[F]&=\frac{1}{m!}\int[d\Phi_m](F|U(t_F,t_0)|\Phi_m)
  (\Phi_m|S) \\
  &+\frac{1}{(m+1)!}\int[d\Phi_{m+1}](F|U(t_F,t_0)|\Phi_{m+1})
  (\Phi_{m+1}|H)\;,
\end{split}
\end{equation}
where $\Phi_m = \{p,f, \hat{c},\hat{s},c,s\}_m$. First we have to
generate the samples
\begin{equation}
\begin{split}
  (\Phi_m|S) &\equiv (\Phi_m|\rho^{(0)}_m) +
  (\Phi_m|\rho_m^{(1)}) + (\Phi_m|\mathbf{I}(t_0)
  +\mathbf{K}(t_0)+\mathbf{P}|\rho_m^{(0)})\;,\\
  (\Phi_{m+1}|H) &\equiv (\Phi_{m+1}|\rho^{(0)}_{m+1}) - \sum_l
  (\Phi_{m+1}|\mathbf{S}_l|\rho_m^{(0)})\Theta(t_l-t_0)\;,
\end{split}
\end{equation}
and then apply the shower evolution operator $U(t_F,t_0)$.
\subsection{Exclusive processes}
The situation is more complex for exclusive processes. In order
to avoid double counting, we have to add inclusive jet functions $F_I$
and we have to modify the subtraction terms
\begin{equation}
\begin{split}
  \bar{\sigma}[F]&=\frac{1}{m!}\int[d\Phi_m](F|U(t_F,t_0)|\Phi_m)
  (\Phi_m|S)F_I(\{\hat{p},\hat{f}\}_m) \\
  &+\frac{1}{(m+1)!}\int[d\Phi_{m+1}](F|U(t_F,t_0)|\Phi_{m+1})
  (\Phi_{m+1}|\tilde{H})F_I(\{p,f\}_{m+1})\;,
\end{split}
\end{equation}
with
\begin{equation}
  (\Phi_{m+1}|\tilde{H}) \equiv (\Phi_{m+1}|\rho_{(m+1)}^{(0)}) -
  \sum_l (\Phi_{m+1}|\mathbf{S}_l|\rho^{(0)}_m)
  \Theta(t_l-t_0)F_I(Q_l(\{p,f\}_{m+1}))\;,
\end{equation}
where the jet function
$F_I(Q_l(\{p,f\}_{m+1})) = F_I(\{\hat{p},\hat{f}\}_m)$ acts
on the underlying Born kinematics. Expanding the evolution operator
now gives
\begin{equation}
\begin{split}
  \bar{\sigma}[F] \approx \sigma^{NLO} + & \int \frac{[d\Phi_m]}{m!}\frac{[d\Phi_{m+1}]}{(m+1)!}
  \int_{t_0}^{t_F}d\tau (F|\Phi_{m+1})(\Phi_{m+1}|\mathcal{H}_I(\tau)|\Phi_m) \\
  &\times(\Phi_m|\rho^{(0)}_m)\left[1-F_I(\{p,f\}_{m+1})\right]F_I(\{\hat{p},\hat{f}\}_m)
  + \mathcal{O}(\alpha_s^2)\;.
\end{split}
\end{equation}
Thus, double counting is removed if $F_I(\{p,f\}_{m+1}) = 1$ for
$F(\{p,f\}_{m+1}) \neq 0$, i.e. when generation cuts are more
inclusive than cuts on the final observable.

\section{Application: $pp\to t\bar{t}j + X$}
The implementation details of the scheme presented in the
previous section in the \helacnlo{} framework \cite{Czakon:2009ss,
*Bevilacqua:2013iha, *vanHameren:2009dr, *Ossola:2007ax}
can be found in Ref.~\cite{Czakon:2015cla}.
We now present results for $pp \to t\bar{t}j + X$  at the LHC.
The NLO QCD corrections have been already presented in Ref.
\cite{Dittmaier:2007wz,*Dittmaier:2008uj,*Melnikov:2010iu,
*Melnikov:2011qx}. First parton shower matched calculations using
the \powheg{} method were presented in Ref.~\cite{Kardos:2011qa,
*Alioli:2011as}, while merging several matched calculations for
different jet multiplicities is discussed in Ref.~\cite{Hoeche:2014qda}.

We consider the LHC at $\sqrt{s}=8$ TeV. The top quark mass is
$m_t=173.5$ GeV. In the shower we set the parton masses of
charm and bottom quarks to $m_c = 1.4$ GeV and $m_b=4.75$ GeV. We use
the MSTW2008NLO PDF~\cite{Martin:2009iq} set in our calculation and
provided it at $\mu_F=1$ GeV to \deductor{}. Renormalization and
factorization scales are set to $\mu_R=\mu_F=\mu_0=m_t$. We use the
anti-$k_T$ jet algorithm~\cite{Cacciari:2008gp} with $\Delta R=1$
and the analysis cuts are $p_T(j_1) > 50$  GeV and $|y(j_1)| < 5$,
while the generation cut is $p_T(j_1) > 30$ GeV. The initial shower
time is chosen to be
\begin{equation}
  e^{-t_0}=\min_{i\ne j}\left\{ \frac{2p_i\cdot p_j}{\mu_T^2 Q^2}
  \right\}\;,
\end{equation}
where the parameter $\mu_T$ allows us to address parton shower
uncertainties. For our central prediction we choose $\mu_T=1$. The
shower evolution is restricted to LC and spin-averaged contributions.
In addition we do not include non-perturbative effects or
top quark decays. The comparison with other Monte Carlo generators is,
therefore, performed at the level of the perturbative evolution.
For the comparison we use \amcnlo{}~\cite{Alwall:2014hca} with the
\mcnlo{} matching in conjunction with \pythia{}8
\cite{Sjostrand:2007gs} ($p_T$ ordered parton shower) and \pythia{}6Q
\cite{Sjostrand:2006za} (virtuality ordered parton shower). For the
\powheg{} matching we use the \powhegbox{}~\cite{Alioli:2010xd}
results together with \pythia{}8.

\begin{center}
\begin{figure}
  \includegraphics[width=0.45\textwidth]{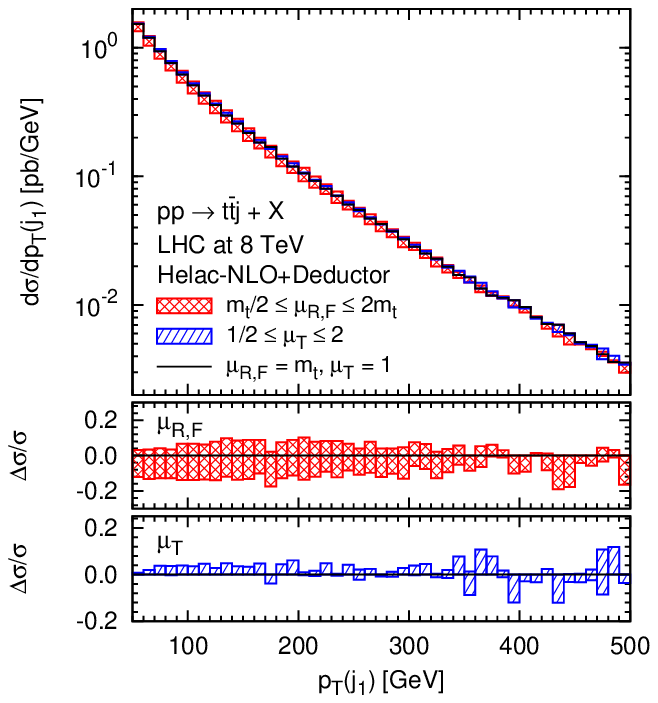}
  \includegraphics[width=0.45\textwidth]{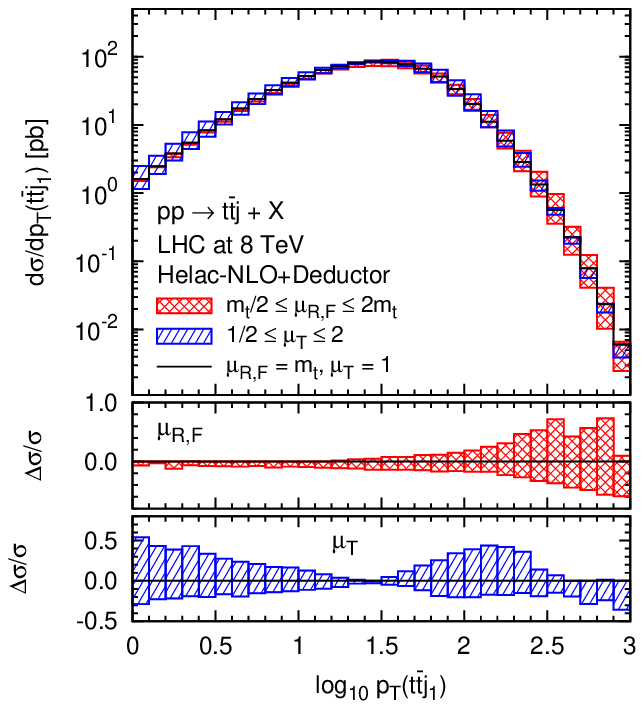}
  \caption{\textit{Differential cross section as a function of the
  transverse momentum of the hardest jet (left panel)
  and of the $t\bar{t}j_1$ system (right panel).}}
  \label{fig1}

  \includegraphics[width=0.45\textwidth]{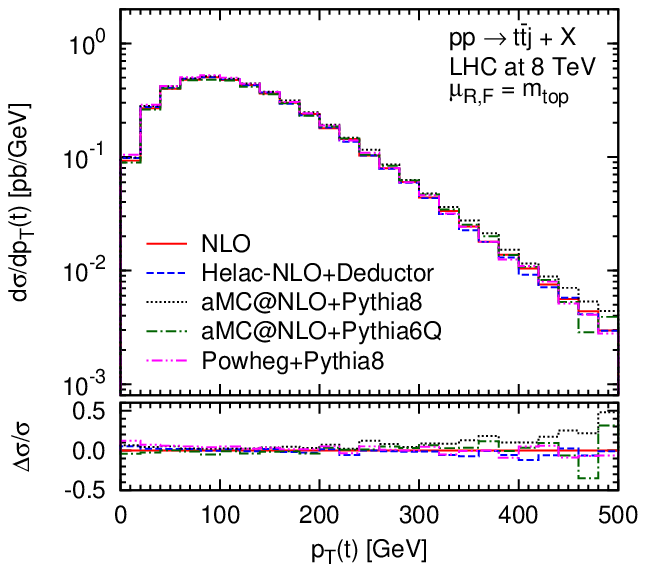}
  \includegraphics[width=0.45\textwidth]{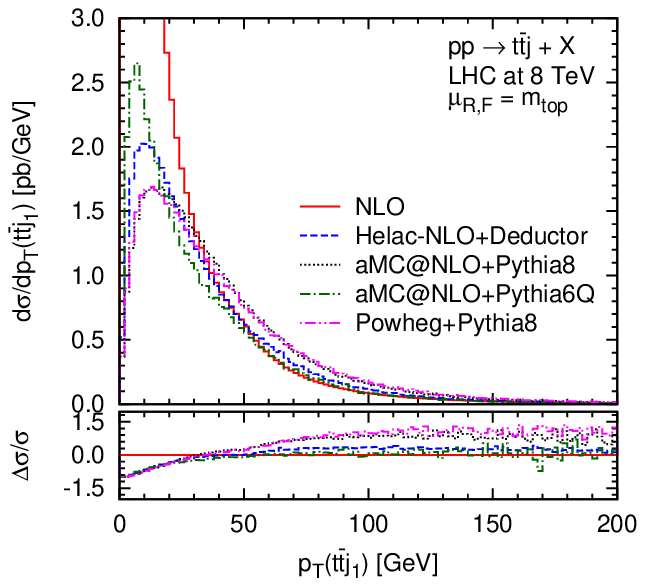}
  \caption{\textit{Differential cross section as a function of the
  transverse momentum of the top quark (left panel)
  and of transverse momentum of the $t\bar{t}j_1$ system (right panel).}}
  \label{fig2}
\end{figure}
\end{center}

In Fig.~\ref{fig1} we address scale and parton shower uncertainties:
$m_t/2 < \mu_0 < 2m_t$ and $1/2 < \mu_T < 2$. The $p_T$ of the
hardest jet shows a flat and reduced scale variation with respect to
both parameters $\mu_0$ and $\mu_T$. This is expected since it is
already an NLO-accurate observable. On the contrary, the $p_T$ of the
$t\bar{t}j_1$ system presents a stronger dependence on the shower
parameter $\mu_T$ in the low $p_T$ regime, where the parton shower
dominates. The scale dependence sets in once the real matrix element
is present, i.e roughly at 30 GeV, and grows rapidly with transverse
momentum.

Fig.~\ref{fig2} illustrates the comparison with other MC generators.
For inclusive observables, like the $p_T$ of the top quark in the left
panel, we do not observe substantial differences between the MC
generators. This is expected because this observable is already
accurate at NLO and serves as a good cross check of our implementation.
For exclusive observables like the $p_T$ of the $t\bar{t}j$ system,
that is shown in the right panel of Fig.~\ref{fig2}, we find a strong
dependence on the initial shower conditions. We observe that
\amcnlopy{} and \powhegpy{} overshoot the
high $p_T$ tail, where one would like to recover the real matrix
element description. On the other hand, \heldeductor{} and \amcnlopyQ{}
preserve the prediction in the high energy tail. The comparison for
other observables can be found in Ref.~\cite{Czakon:2015cla}.
%
\section{Summary and Outlook}
We presented the NLO matching scheme for the
Nagy-Soper parton shower in the spirit of the \mcnlo{} method.
We studied the production of $t\bar{t}j$ at the LHC using the
\heldeductor{} framework and compared it to the other MC generators.
We want to stress that the current accuracy of \heldeductor{} is only
LC and spin-averaged, i.e  the same as for other MC programs.
However, this comparison is an important validation of our
implementation. Differences, present in exclusive observables, can be
traced back, for example, to the choice of the initial shower
conditions.

In the future we want to extend the matching scheme implemented in
\helacnlo{} to include full colour and spin correlations. Nevertheless,
those effects have to be first added to \deductor{}.

\section*{Acknowledgments}
The author acknowledges support by the DFG under Grant No. WO 1900/1-1
(\textit{"Signals and Backgrounds Beyond Leading Order. Phenomenological
studies for the LHC"})

\end{document}